\documentclass[epj,nopacs,floatfix]{svjour}
\usepackage{graphicx}
\usepackage{bm}
\usepackage{amssymb}
\usepackage{cite}

\newcommand{\pt}{$p_T$}
\begin{document}
\hyphenation{CH-A-MP}
\hyphenation{CH-A-MPs}

\title{Searching for high speed long-lived charged massive particles at the LHC}
\author{Jie Chen\thanks{\email{jchen@fnal.gov}} \and Todd Adams\thanks{\email{tadams@hep.fsu.edu}}
}                   
\institute{Department of Physics, Florida State University, Tallahassee, FL 32306, USA
}

\date{Received: date / Revised version: date}
\abstract{
The conventional way to search for long-lived CHArged Massive Particles 
(CHAMPs) is to identify slow (small $\beta$) tracks 
using delayed time of flight and high ionization energy loss. 
But at the $7-14$ TeV center of mass energy of the LHC, a CHAMP may
be highly boosted (high $\beta$) and therefore look more like a 
minimum ionizing particle, while for high momentum muons 
($\gtrsim$500 GeV/c) the radiative effect dominates energy 
deposition.  This suggests a new strategy to search for CHAMPs at 
the LHC.  Using energy deposition from different detector 
components, we construct a boosted decision tree discriminant to 
separate high momentum CHAMPs from high momentum muons.  This method 
increases substantially the CHAMP discovery potential and it can be used 
to distinguish possible di-CHAMP or CHAMP-muon resonance models from 
di-muon resonance models.  We illustrate the new method using a mGMSB 
model and a recently proposed di-CHAMP model and we give updated 
CHAMP mass limits for these two models using the results from a recent CDF CHAMP search.
}
\maketitle

\section{\label{sec:intro}Introduction}
Many extensions of the standard model (SM) suggest the existence 
of a CHArged Massive Particle (CHAMP) that is 
long-lived~\cite{GMSB,splitSUSY,UEDLKP,MSSM,mSugra,dichamp,leptosusy,coannih}.  
The long lifetime is caused by limited phase space due to small mass 
splittings or by small couplings.
If the CHAMP escapes a particle detector before decaying
it can be regarded as stable.  CHAMP searches play an important 
role in constraining SUSY models (see, for
example, Ref.~\cite{SUSYwoprejudice}). Therefore, there is interest in pursuing these
searches using early LHC data. 

A number of searches for weakly interacting CHAMPs were performed at LEPII~\cite{LEPhscpL3, LEPhscpOPAL, LEPhscpALEPH, LEPhscpDELPHI}, yielding 95$\%$ C.L. lower mass 
limits of $\sim$100 GeV/$c^2$ for a stable supersymmetric partner of SM leptons~\cite{LEPCombine}.  Tevatron searches for
CHAMPs have been carried out recently by CDF~\cite{cdfchamp}
and D0~\cite{d0champ}.
The conventional method to detect CHAMPs is to search for slow 
moving and/or highly ionizing particles~\cite{Reporthscp}. 
This technique has been used for the LEP and 
Tevatron searches, where the CHAMPs are expected to be slow due to the low 
center of mass energy and the expected high CHAMP mass ($>$100 GeV/$c^2$).

At the energy of the LHC ($\sqrt{s} = 7-14$ TeV), however, CHAMPs can be highly boosted. 
Therefore, they behave more like minimum ionizing 
particles (MIPs), and consequently, the conventional search
technique may lead to significant signal losses. 
Current experimental efforts at the LHC~\cite{hscpcmspas,atlashscp} 
focus on slow moving CHAMPs, separating them from muons using their larger 
ionization energy loss and time of flight delays. In this paper, 
we propose a way to search for high speed CHAMPs. 
We use the difference in energy deposition 
in the sub-detectors, as high speed (high momentum, $p$) CHAMPs behave 
like MIPs while high $p$ muons tend to deposit more energy due to 
radiation effects.  The goal is to increase the discovery potential
of the LHC experiments using early data.

\section{\label{sec:models}Models Under Investigation}

\begin{figure}
 \unitlength1cm
 \begin{picture}(8.0,8.0)
  \put(0.0,4.0) {\includegraphics[width=4cm]{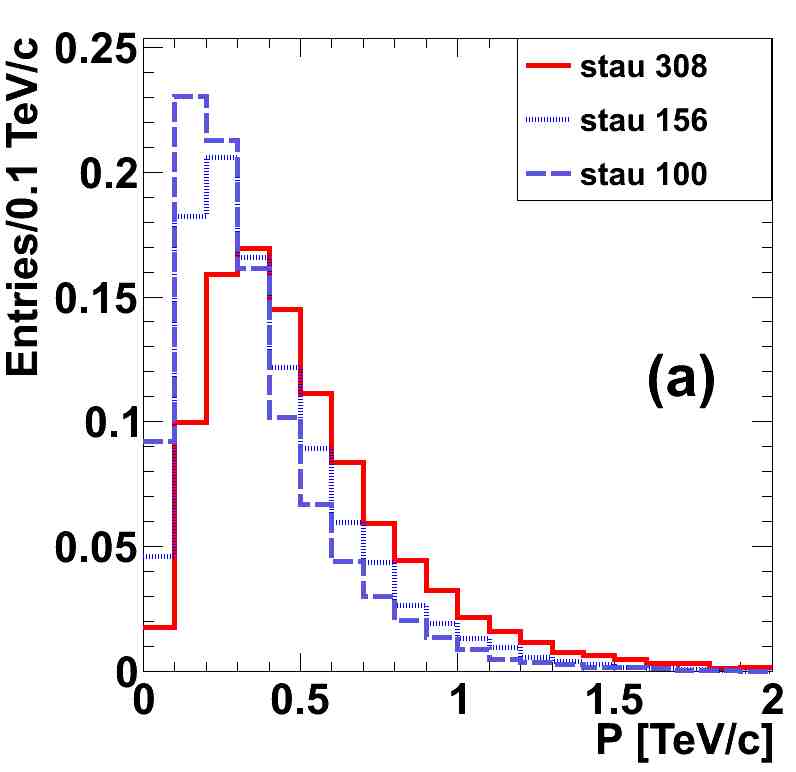}}
  \put(4.0,4.0) {\includegraphics[width=4cm]{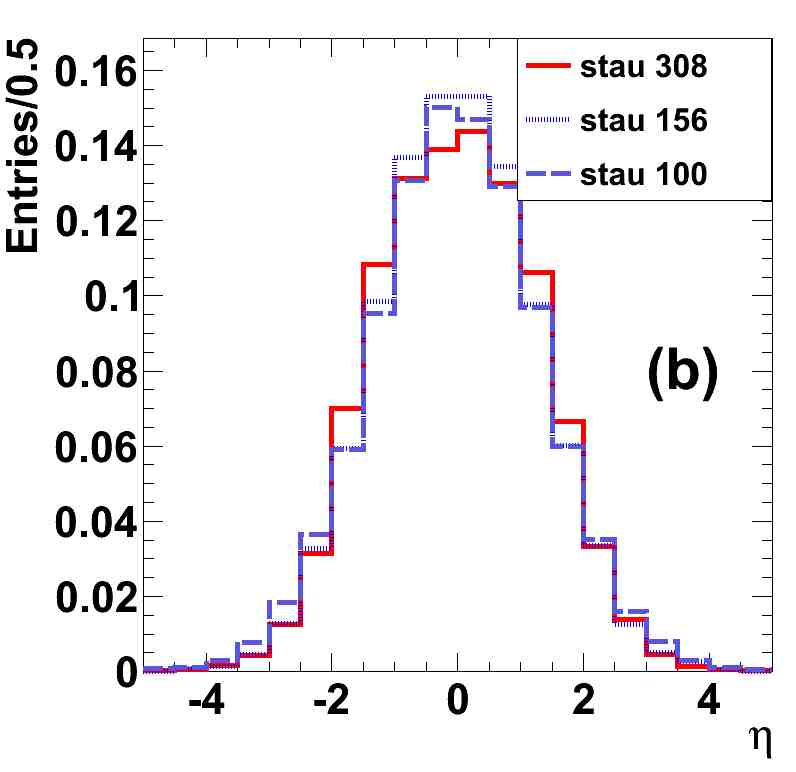}}
  \put(0.0,0.0) {\includegraphics[width=4cm]{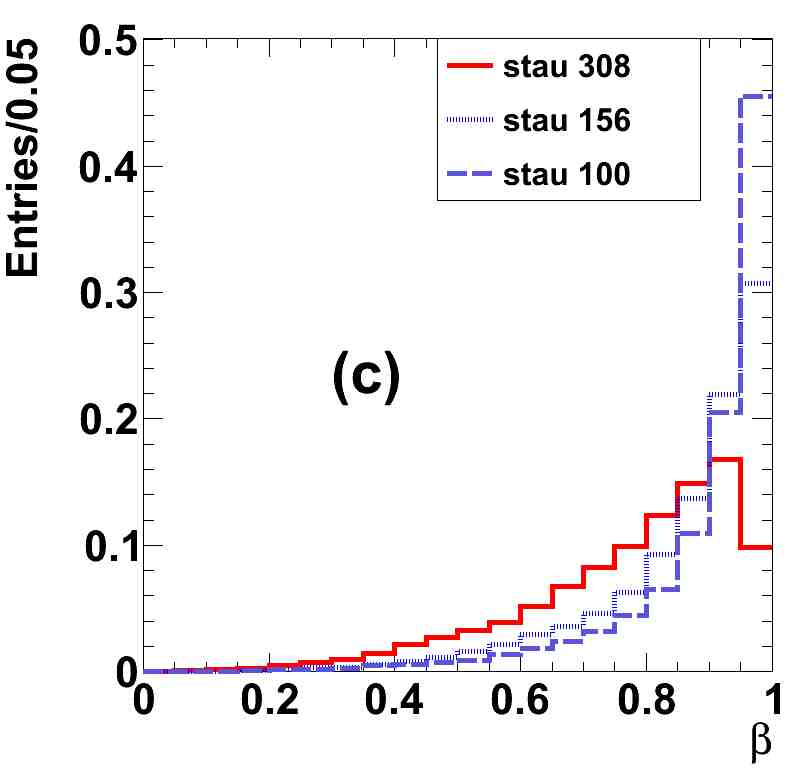}}
  \put(4.0,0.0) {\includegraphics[width=4cm]{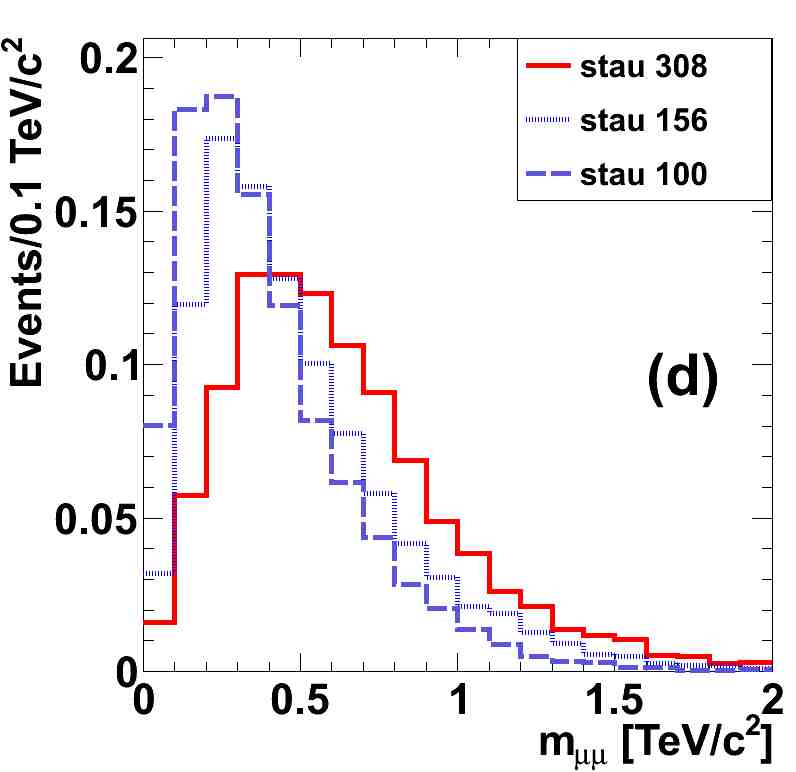}}
  \end{picture}
 \caption{\label{fig:staugendist} Distributions for some kinematic quantities
    for the SPS7 benchmark scenario at stau masses of 100,
    156, and 308 GeV/$c^2$ for: (a) stau $p$,
    (b) stau $\eta$, (c) stau $\beta$, (d) di-stau invariant mass 
    (assuming track has zero mass).  
    All distributions are normalized to unit area.}
\end{figure}

We investigate two models for this study:
the minimal gauge mediated supersymmetry breaking (mGMSB) 
model\cite{GMSB} 
and a resonant di-CHAMP model~\cite{dichamp}.  In
all simulations, we use $\sqrt{s} = 10$ TeV for the LHC and
$\sqrt{s} = 2$ TeV for the Tevatron.
The mGMSB model we explore corresponds to the SPS7 benchmark scenario proposed in~\cite{SPS7}. In this model, the stau ($\tilde{\tau}$) is the next-to-lightest
supersymmetric particle (NLSP).  We vary $\Lambda$ from 
31 TeV to 100 TeV, with fixed parameters 
$N_{mes}$ = 3, 
$\tan\beta$ = 15, $\mu > 0$, $C_{{grav}}$ = 10000 and 
$M_{{mes}}/\Lambda$ = 2. 
The large value of $C_{{grav}}$ results in a long-lived stau, while
$\Lambda = 31-100$ TeV gives a stau mass of 
100 GeV/$c^2$ to 308 GeV/$c^2$. 
We produce the mGMSB particle mass spectrum and the decay table 
with the program {\sc isasugra}~\cite{isasugra}, which is used as
input to {\sc pythia 6.4}~\cite{pythia} with all SUSY processes
enabled.  Figure~\ref{fig:staugendist} shows the
distribution for some kinematic quantities for three different values of the stau mass within this benchmark scenario, at an LHC energy of 10 TeV. 

\begin{figure}
 \unitlength1cm
 \begin{picture}(8.0,8.0)
  \put(0.0,4.0){\includegraphics[width=4cm]{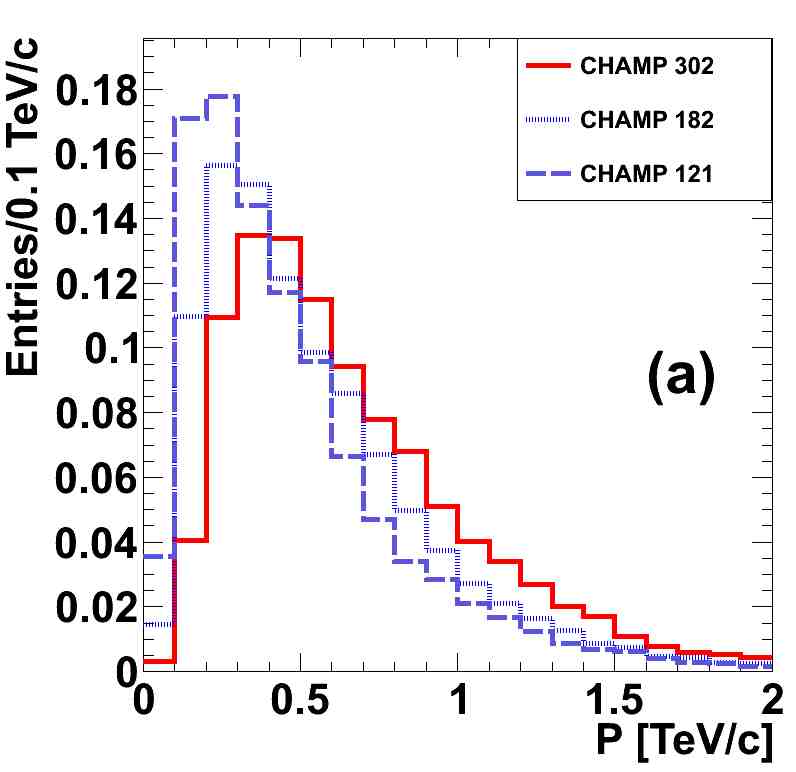}}
  \put(4.0,4.0){\includegraphics[width=4cm]{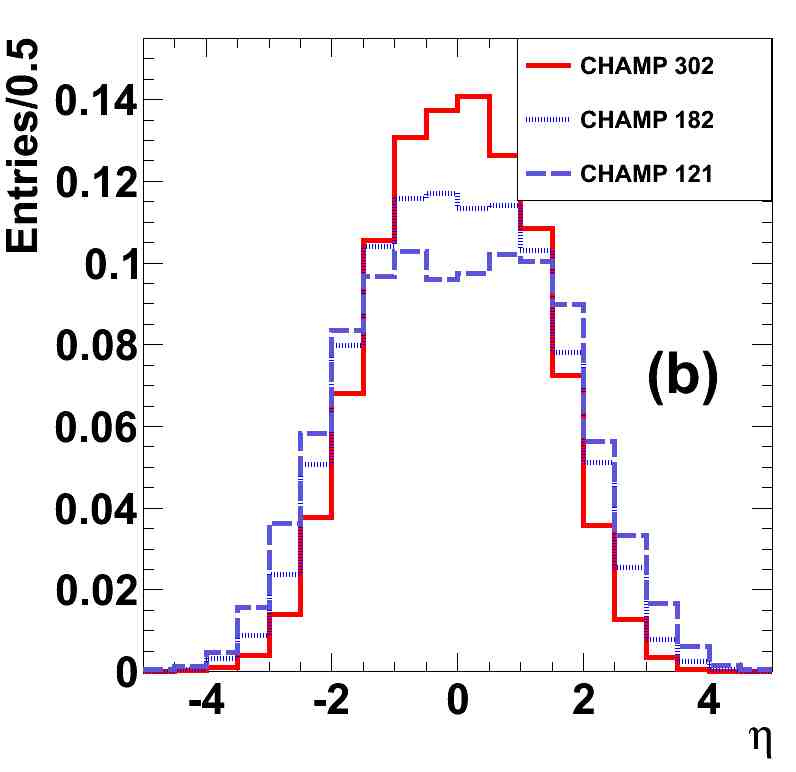}}
  \put(0.0,0.0){\includegraphics[width=4cm]{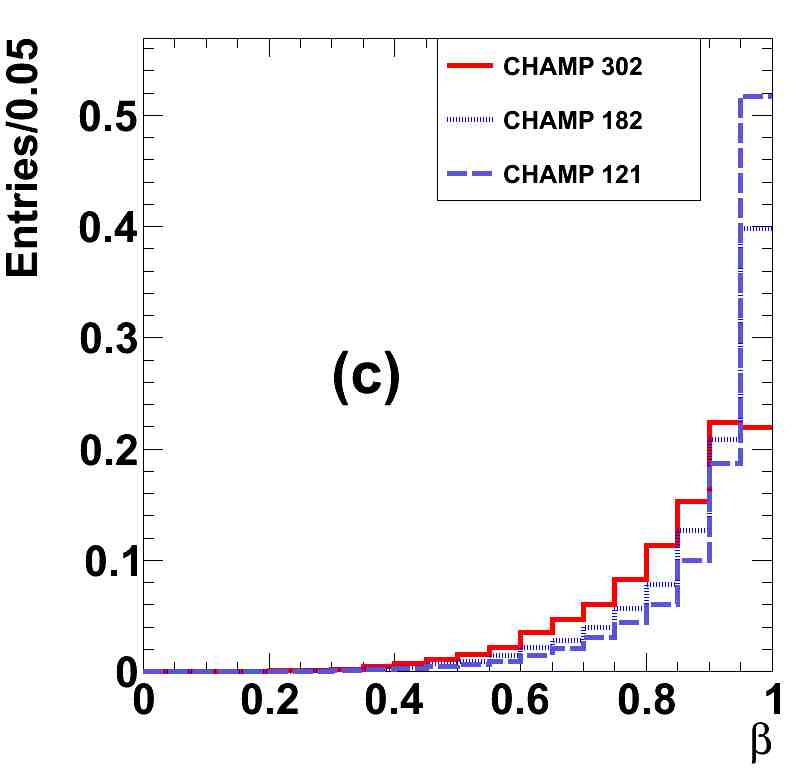}}
  \put(4.0,0.0){\includegraphics[width=4cm]{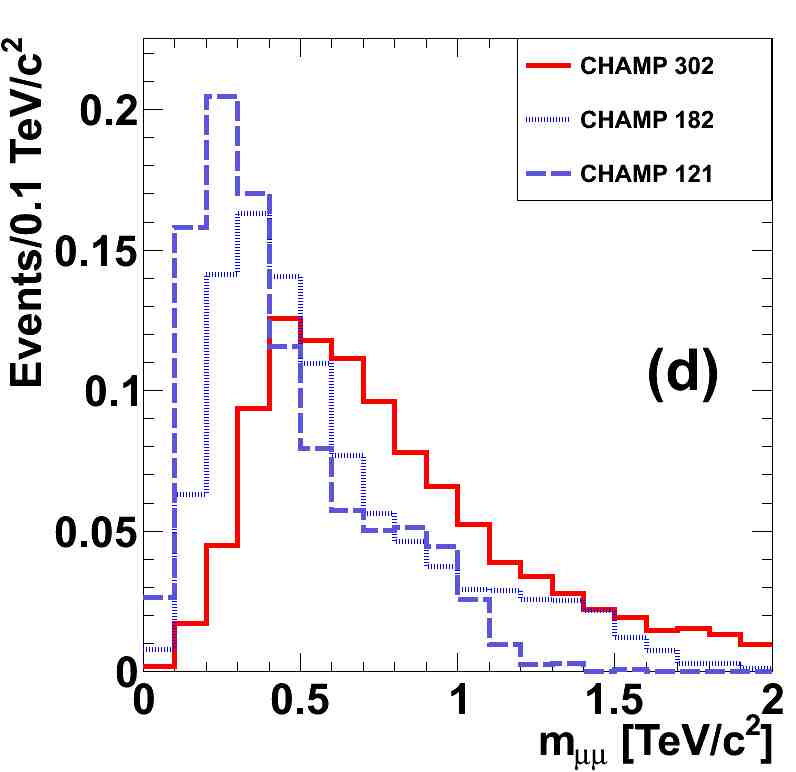}}
 \end{picture}
 \caption{\label{fig:dichampgendist} Distributions for some kinematic quantities
    for the di-CHAMP model with CHAMP mass 121, 182 and 302 GeV/$c^2$
    for: (a) CHAMP $p$, (b) CHAMP $\eta$, (c) CHAMP $\beta$, 
    (d) di-CHAMP invariant mass (assuming track has zero mass).
    All distributions are normalized to unit area.}
\end{figure}

Some recent models predict di-CHAMP~\cite{dichamp}, 
or CHAMP-muon~\cite{leptosusy} resonances in which  the CHAMPs 
are relatively light compared to the resonances, and thus obtain a large boost. 
We generate di-CHAMP events using {\sc CompHEP} 4.5.1~\cite{Comphep} with a
model file provided by Kilic, Okui and Sundrum
based on Ref.~\cite{dichamp}. The {\sc CompHEP} events are hadronized using
the program {\sc Pythia}. We vary 
the $M_{\tilde \rho}$ parameter in the di-CHAMP model from 
1.0 TeV/$c^2$ to 2.5 TeV/$c^2$, 
yielding CHAMP ($\tilde K$) masses between 121 GeV/$c^2$ and 302 GeV/$c^2$.  The $\tilde K$ to $\tilde \rho$ mass ratio is 0.12 in the model used and the factorization scale is set to the invariant mass of the di-CHAMP system. 
Figure~\ref{fig:dichampgendist} shows the
distribution for some kinematic quantities for this model at 10 TeV. 

Both models predict events  that may contain two CHAMPs.  The di-CHAMP 
invariant mass distributions are shown in Figs.~\ref{fig:staugendist}
and~\ref{fig:dichampgendist}.  Because high $\beta$ CHAMPs
resemble muons in their experimental signatures (for example, they are deeply
penetrating and have a short travel time to the detectors), these events could 
appear in searches for
high mass di-$\mu$ resonances (as predicted by
$Z^\prime$ models~\cite{Littlehiggs,RSgravitonZp,nuzprime,Stzprime})
or in the high mass tails of di-$\mu$ distributions (as in contact interaction
models~\cite{compositeness,cdfcompositeseach}).

\begin{figure}
 \includegraphics[width=8cm]{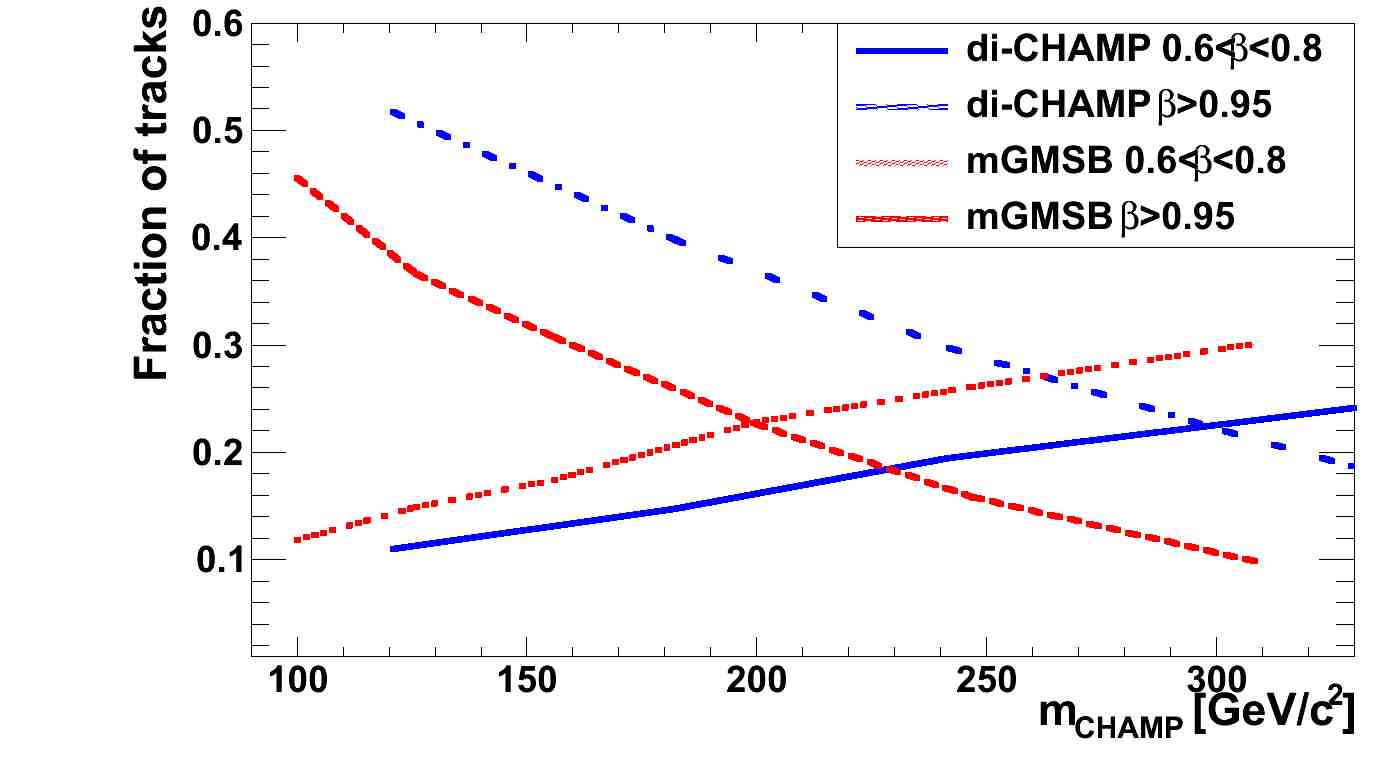}
 \caption{\label{fig:numhighbeta} The fraction of tracks that 
    fall within $\beta > 0.95$ and $0.6 < \beta < 0.8$
    for the mGMSB stau model (red) and the 
    di-CHAMP model (blue) at 10 TeV LHC energy.}
\end{figure}

We define $\beta > 0.95$ as the high momentum (muon-like) region 
and $0.6 < \beta < 0.8$ as the traditional CHAMP region.  
The 0.6-0.8 $\beta$ range is the approximate range of the planned 
LHC experimental
searches.  Both ATLAS and CMS have acceptance for $\beta < 0.6$, 
but the efficiency falls rapidly~\cite{hscpcmspas,atlashscp} below
this value. 
Figure~\ref{fig:numhighbeta} shows the fraction of events
in the two regions for the mGMSB
and di-CHAMP models at 10 TeV.  
For both models, more 
events are produced in the high momentum region than in the
traditional search region for CHAMPs of lower mass.  Given the
expected higher trigger and reconstruction efficiency for the high 
momentum region than for the traditional CHAMP search region, we
would expect to  observe more high momentum than low momentum 
CHAMPs when their masses are low.  

Previous experimental results from LEP and the Tevatron limit the
interesting phase space for high momentum CHAMPs at the LHC.
CDF has set a limit of 10 fb on the cross section to produce
a single, weakly interacting CHAMP.  To apply this limit to
our benchmark models, we must account for the fact that some
events may have two CHAMP candidates.
Therefore, we calculate the acceptance, $\alpha$,
for an event to pass the CDF slowly moving minimum ionizing particles trigger
 criteria 
($p_T > 40$ GeV/c, $|\eta| <  0.7$, $0.4 < \beta < 0.9$, 
and isolation) given one or two chances:
\begin{equation}
\alpha = \frac{N_{1}+ N_{2}\times \frac{\epsilon_{2}}{\epsilon_{1}}}{N_{tot}},
\end{equation}
where $N_{tot}$ is the total number of events generated, 
$N_1$ ($N_2$) is the number of events with one (two) CHAMPs within
the trigger criteria, $\epsilon_1$ is the efficiency given by
Ref.~\cite{cdfchamp} to find a CHAMP given one chance, and 
$\epsilon_2$ is the efficiency to find a CHAMP given two 
chances~\cite{privatethomas}.  

\begin{figure}
 \unitlength1cm
 \begin{picture}(8.0,6.0)
  \put(-0.0,0.0){\includegraphics[width=8.5cm]{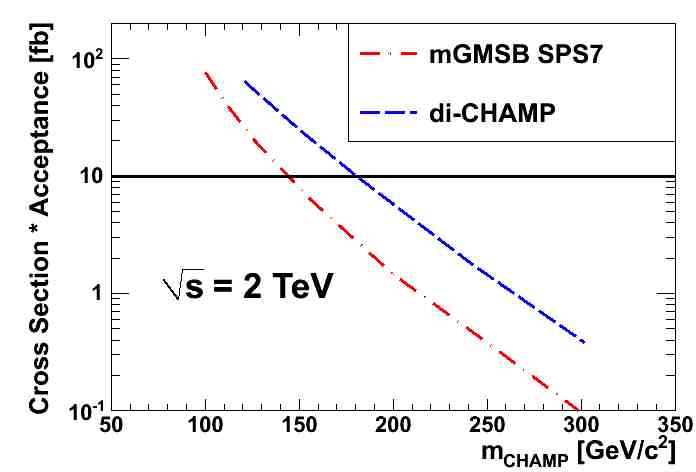}}
 \end{picture}
 \caption{\label{fig:cdflimit} Cross section times acceptance
    for the mGMSB SPS7 benchmark scenario (red) and di-CHAMP model 
    (blue) at $\sqrt{s} = 2$ TeV.  The horizontal black line shows the CDF
    limit on single, weakly-interacting CHAMPs~\cite{cdfchamp}.}
\end{figure}

Figure~\ref{fig:cdflimit} shows the cross section times 
acceptance for these models as a function of the CHAMP mass.  
We interpret this plot as excluding stau masses below $\sim$140 GeV/$c^2$ 
and di-CHAMP resonances with CHAMP masses below $\sim$180 GeV/$c^2$.
Tables~\ref{tab:sps7scan} and~\ref{tab:dichampscan} list the 
cross sections for the two models, and the corresponding CDF acceptance 
at $\sqrt{s} = 2$ TeV.
Based on these exclusion limits, we 
select the stau mass = 156 GeV/$c^2$ in the mGMSB model and the $\tilde K$ 
mass = 182 GeV/$c^2$ in the di-CHAMP model as our benchmark signals 
(labeled stau156 and di-CHAMP182 in the following). 

\begin{table}
\centering \caption{Cascade and pair production cross sections at Tevatron and LHC energies for the SPS7 benchmark scenario in mGMSB.  The last column shows the acceptance for an event to have at least one CHAMP within the CDF analysis selection for cascade production at the Tevatron energy.}

\begin{tabular}{c|c|c|c|c|c}
\hline
 & \multicolumn{2}{c|}{$\sigma(\sqrt{s}=2$ TeV)} & \multicolumn{2}{c|}{$\sigma(\sqrt{s}=10$ TeV)} & Acceptance \\ 
mass & cascade & pair & cascade & pair & for CDF \\ 
(GeV/$c^2$) & ($fb$) & ($fb$) & ($fb$) & ($fb$) & analysis\\ 
\hline 
100 & 113   & 10.6  & 4716 & 57   & 0.68 \\
112 &  58   &  7.5  & 2433 & 34  & 0.69 \\
126 &  28   &  4.8  & 1212 & 21   & 0.74 \\
156 &  7.7  &  1.7  &  320 & 11   & 0.82 \\
200 &  1.6  &  0.5  &   88 &  4   & 0.90 \\
247 &  0.43 &  0.2  &   29 &  2   & 0.96 \\
308 &  0.08 &  0.03 &    9 &  0.8 & 0.96 \\
\hline
\end{tabular}
\label{tab:sps7scan}
\end{table}

\begin{table}
\centering \caption{Production cross sections at Tevatron and LHC energies for the di-CHAMP model. The last column shows the acceptance for an event to have at least one CHAMP within the CDF analysis selection at 2 TeV Tevatron energy.}
\begin{tabular}{c|c|c|c}
\hline
mass &$\sigma (\sqrt{s} = 2$ TeV)&$\sigma (\sqrt{s} = 10$ TeV)& Acceptance\\  
(GeV/$c^2$)&($fb$)&($fb$)&CDF \\ 
\hline
121 & 90  & 845   & 0.71\\
151 & 30  & 360   & 0.80\\
182 & 11  & 175   & 0.87\\
242 & 2.0 &  53 & 0.93\\
302 & 0.4 &  20 & 0.96\\
\hline
\end{tabular}
\label{tab:dichampscan}
\end{table}

\section{\label{sec:event}Event Generation and Selection }

In addition to the signal (see Sec.~\ref{sec:models}), we
model multiple sources of SM (muon) backgrounds ($W$, $Z/\gamma^*$, 
multijet, $t\bar{t}$, and diboson) using {\sc Pythia 6.4}.  All samples 
are generated using  the CTEQ 5L parton distribution 
functions (PDFs)~\cite{cteq5l}.  
We will use CMS as
our reference LHC detector for this study.  However, the principles
(modified appropriately) are also applicable to ATLAS.

To select CHAMP candidate events, we require at least one high 
momentum ($p >  500$ GeV/c) muon-like track  within the
$|\eta| < 1.479$ region.  The 500 GeV/c momentum requirement corresponds 
to $\beta > $ 0.95 for the stau156 and 
$\beta >$ 0.94 for di-CHAMP182, which are fast enough to appear similar to muons (for
masses at the upper edge of the range considered in
these models, this requirement corresponds to $\beta \gtrsim$ 0.85). 
Moreover, this ensures that background muons are highly boosted and their energy loss is 
dominated by radiation.  The muon critical energy occurs
at a muon momentum of 317 GeV/c in copper, 170 GeV/c in lead tungstate and 581 GeV/c in silicon~\cite{PDG08}.
The $|\eta| < 1.479$ selection corresponds to the barrel part of the 
CMS electromagnetic calorimeter (Ecal)~\cite{cmsdetectorpaper}, 
where the energy measurement is most accurate.  
The $\eta$ requirement also 
helps reject high momentum muon background in the forward region.
In addition, we require a second muon-like track with momentum 
$p > 100$ GeV/c and $p_T  > 20$ GeV/c in the $|\eta| < 2.5$ region to reduce
the $W$+(jets) background. 
Table~\ref{tab:mcsample} shows the cross sections for the signal and
background processes after the leading and second leading CHAMP/muon
requirements.  

After these simple selection criteria, the background level is 
already much smaller than, or similar to, the expected signal for 
the stau156 and di-CHAMP182 benchmark models. 
Therefore, even a simple counting experiment may be sensitive to
a CHAMP signal.  However, due to PDF uncertainties and the fact that 
{\sc Pythia} is only a tree level generator, the background assumed 
here may be underestimated. The 
effect of a larger background will be discussed later.
In any case, the energy deposition in the detector can be used to 
separate further signal from background. 

\begin{table}
\centering \caption{Cross sections for signal and background samples before
selection and after leading and second leading candidate selections.  The
$W/Z$+(jets) and multijet backgrounds have generator level cuts of
30 GeV/c and 215 GeV/c on the leading jet \pt, respectively.}
\begin{tabular}{lccc}
\hline
                     &            & $\geq$1 high p $\mu$ & 2nd  $\mu$\\ 
Process              & $\sigma (pb)$ & $\sigma (pb)$ & $\sigma (pb)$ \\ \hline
$W$+(jets)                &     75600     &     0.3       &      0.002   \\ 
$Z/\gamma^{\ast}$+(jets)                &      7240     &     0.09      &      0.03  \\ 
$t\bar t$            &       234     &     0.03      &      0.003  \\ 
$WZ$,$WW$,$ZZ$             &      69.4     &     0.005     &      0.001   \\ 
multijet                  &     20100     &   $<$0.01     &   $<$0.01  \\ 
\hline
Total Background  &    103000     &     0.43      &      0.04  \\
\hline
stau156              &     0.32      &     0.09      &      0.08 \\ 
di-CHAMP182          &     0.18      &     0.05      &      0.05 \\ 
\hline
\end{tabular}
\label{tab:mcsample}
\end{table}

\section{\label{sec:sim} Detector Simulation}

\begin{figure}
 \unitlength1cm
 \begin{picture}(8.0,4.0)
  \put(-0.3,0.0){\includegraphics[width=9.6cm]{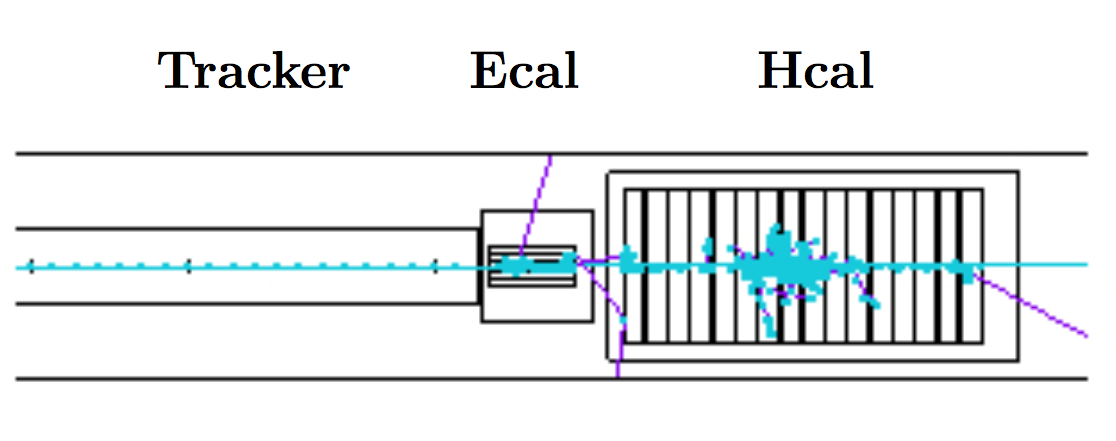}}
 \end{picture}
\caption{\label{fig:cmssimplesim} {\sc geant4} simulation: a muon track with $p$ = 650 GeV/c passing through a simplified model of the CMS detector.}
\end{figure}

We use 
{\sc Geant4}~\cite{geant4} to simulate a simplified model of the CMS 
detector~\cite{cmsdetectorpaper},  shown in Fig.~\ref{fig:cmssimplesim}. 
In our model, the tracker is made of 20 layers of 300 $\mu$m silicon,  the 
electromagnetic calorimeter is a 5 by 5 array of PbWO$_4$ crystal with 
the same dimensions as a CMS Ecal crystal, and the hadron calorimeter (Hcal)
is composed by 30 layers of brass with
transverse size similar to that of a CMS Hcal tower.
No magnetic field is included in our 
simulation, as we expect high momentum tracks to be rather 
straight in the detector.

The energy loss per path length (dE/dx) for a muon or a CHAMP in the tracker is calculated using the truncated mean~\cite{TIFTracker}  estimator in which the arithmetic mean is computed after truncation of the highest 40\% of the charge samples for each track.  We take the Ecal energy to be the energy deposited in the crystal traversed by the CHAMP or muon. This was chosen to minimize the possible energy contribution from nearby charged or neutral particles and to avoid the need to simulate the shower shape in a magnetic field.  
The Hcal energy is computed by summing the energy measurements for all layers of Hcal, which  is roughly the same as the single tower energy defined in the CMS detector. 

To better account for the detector
response, the {\sc Geant4} simulated energy is smeared with the CMS detector 
energy resolution functions~\cite{cmsdetectorpaper}.   For Hcal it is 
${\sigma}/{E}=120\%/\sqrt{E}+6.9\%$~\cite{CMSptdr1}. The Ecal 
energy is smeared using 
$({\sigma}/{E})^2=({2.8\%}/{\sqrt{E}})^2+({0.12}/{E})^2+(0.30\%)^2$.
For tracker dE/dx, we do not smear the energy because the simulated 
energy resolution is approximately the same as that of the reconstructed energy. 
This is due to the very high signal to noise ratio for the 
CMS tracker readout electronics~\cite{cmsdetectorpaper}.

\begin{figure}
 \unitlength1cm
 \begin{picture}(8.0,8.0)
  \put(2.0,4.0){\includegraphics[width=4cm]{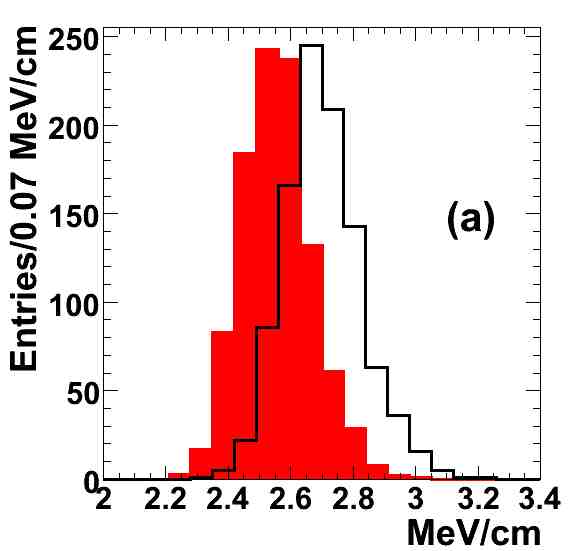}}
  \put(0.0,0.0){\includegraphics[width=4cm]{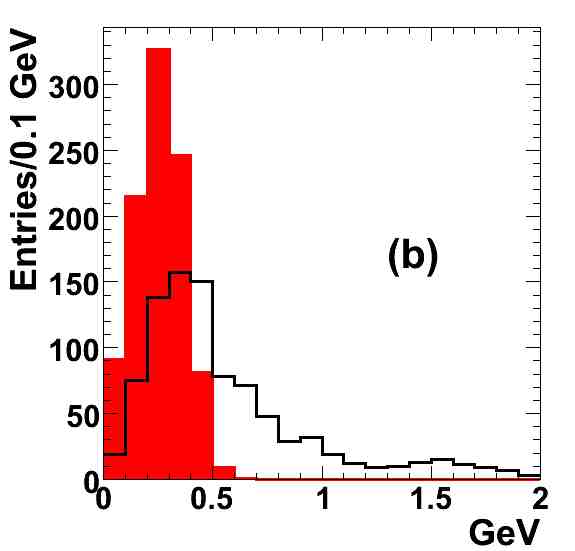}}
  \put(4.0,0.0){\includegraphics[width=4cm]{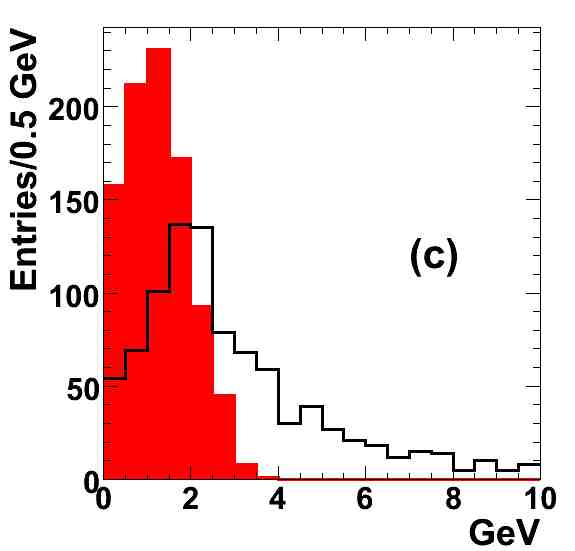}}
 \end{picture}
\caption{\label{fig:energydis} Distributions of (a) dE/dx in the 
         Tracker, (b) energy deposition in Ecal, and (c) energy
         deposition in Hcal.  The high momentum stau156 signal and the background are shown
         with a solid and an empty histogram respectively.}
\end{figure}

As noted in Sec~\ref{sec:event}, the main background is $Z/\gamma^\ast$+(jets) after our simple selections. 
We save all high momentum tracks from all 
background or signal events and create background samples by mixing
events according to their cross sections.  
The muon/CHAMP tracks are then passed to our simplified CMS {\sc geant4} 
simulation and we obtain the energy deposition for each track.

The results of our  simulation and energy smearing indicate that the distributions of CHAMP energies in Ecal and Hcal should be Gaussian (see Fig.~\ref{fig:energydis}). This is due to the fact that CHAMPs primarily lose energy through ionization.  For the thick material in Ecal and Hcal, the Landau distribution becomes Gaussian. The high speed muons, however, exhibit long tails in dE/dx distributions, corresponding to radiative energy losses. 
For the tracker, the dE/dx distribution is Gaussian in both cases, as expected for
the truncated mean estimator~\cite{TIFTracker}.

\begin{figure}
\includegraphics[width=9cm]{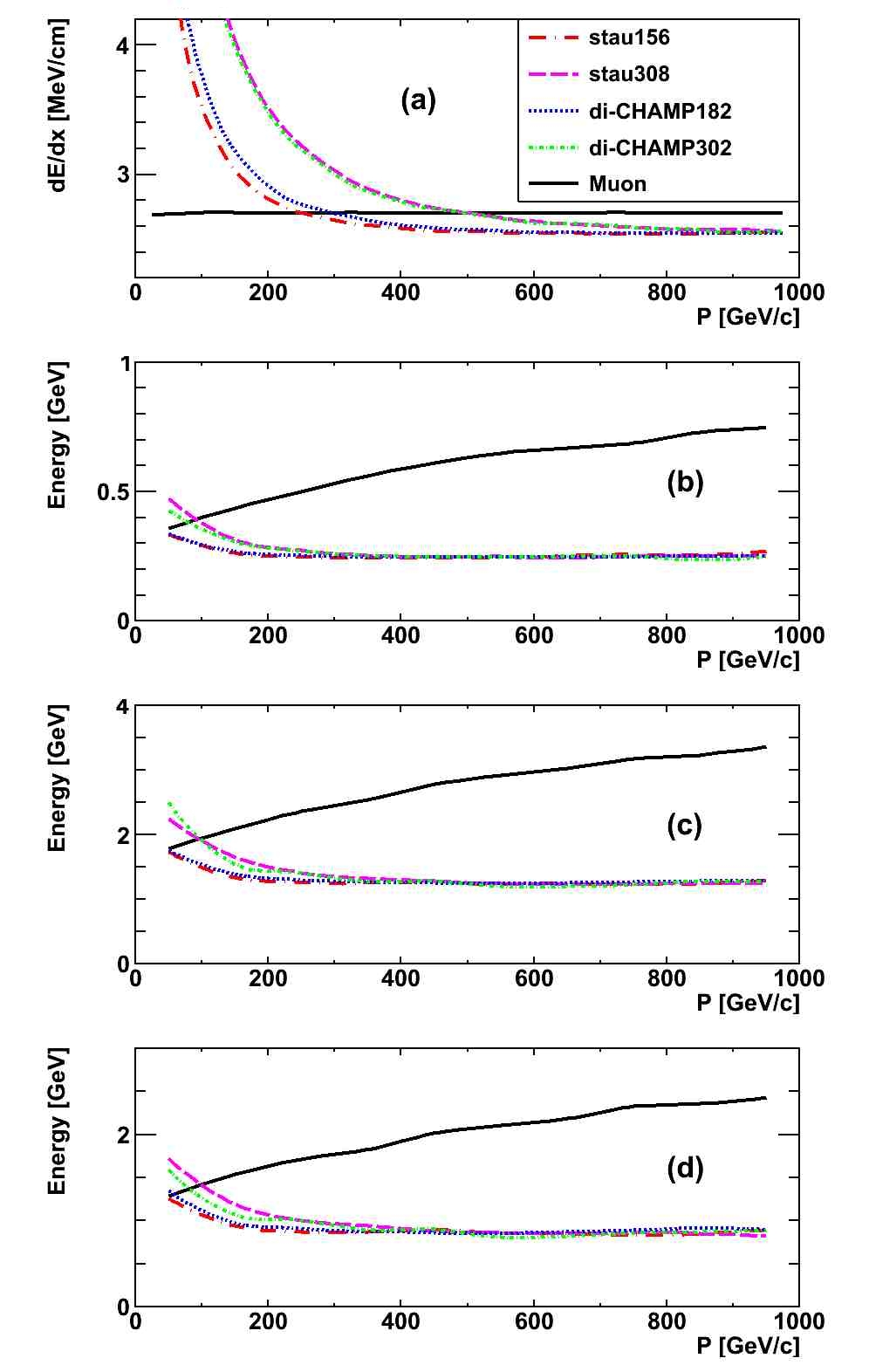}
\caption{\label{fig:evsp} The average (a) Tracker dE/dx, (b) energy deposition in Ecal, (c) energy deposition in Hcal, and (d) difference between Hcal and Ecal in energy deposition,  as a function of CHAMP/muon momentum for four different CHAMP model/mass points and background muons. Further analysis in this paper will concentrate on the high momentum ($\gtrsim$ 500 GeV/c) region.  }
\end{figure}

In Fig.~\ref{fig:evsp}, we show the average values of Tracker dE/dx,  
energy deposition in Ecal, energy deposition in Hcal, and energy 
deposition difference between Hcal and Ecal as a function of momentum, 
for two CHAMP models and two mass points. 
The low momentum region ($\lesssim$ 500 GeV/c, corresponding to 
low $\beta$s, which is the focus of conventional CHAMP searches) 
shows the high ionization energy loss, which is inversely 
proportional to $\beta^2$~\cite{PDG08}. 
We make an important observation: the average energy deposition in Ecal and Hcal in the high momentum region ($\gtrsim$ 500 GeV/c) are almost identical for the four model/mass benchmarks  considered. 
Note that the muon energy deposition increases with momentum, as expected from the increased radiative contributions, while the Tracker dE/dx of muons is evenly distributed  in approximately 
the full momentum range, mainly due to the fact that here dE/dx is calculated with 40\% truncation, which does not  take into account high energy (radiative) hits.

Given the important observation that the energy depositions in these detectors is
largely model
and mass independent in the high $p$ region (Fig.~\ref{fig:evsp}), we can use
the energies to develop a new method to separate CHAMPs from muons in a 
manner that is analogous to  standard particle identification that separates
electrons and photons from hadrons and jets~\cite{d0photon}.  In the following
section we use a multivariate method to enhance the discrimination between muons
and CHAMPs.

\section{\label{sec:nn} Multivariate Analysis}

The direct use of the energy depositions, whose distributions are shown in 
Fig.~\ref{fig:energydis}, can only provide limited separation 
between CHAMPs and muons passing our basic selection.  However, it is
well established that better separation between distributions can often be obtained by combining
well understood variables into a multivariate discriminant.
To that end, we use the boosted decision tree (BDT) method provided in
the Toolkit for Multivariate 
Data Analysis (TMVA)~\cite{TMVA} to compute discriminants that take 
the three energy depositions as inputs.

\begin{figure}
\includegraphics[width=8cm]{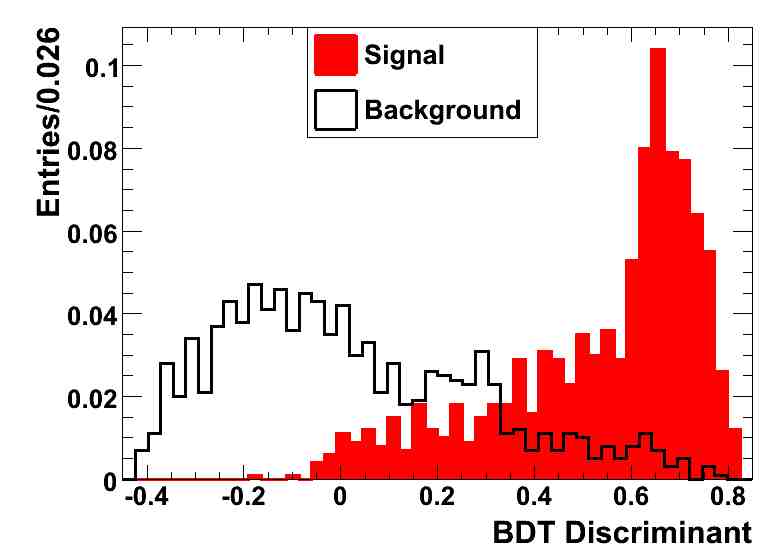}
\caption{\label{fig:nnoutput} Distributions of the BDT discriminant for
   stau156 signal (solid) and background (empty). }
\end{figure}

Figure~\ref{fig:nnoutput} shows the BDT discriminant distribution 
for the stau156 model and backgrounds after our simple selection.  We see very 
clear separation between high speed CHAMPs and background muons.
The di-CHAMP182 model has a behavior very similar to the one of stau156 model. 
We estimate the significance of the signal using
\begin{equation}
 S_{cL}=\sqrt{2((S+B)\ln(1+\frac{S}{B})-S)},
\end{equation}
where S and B are the number of signal and background 
events~\cite{CMSptdr2} that remain after a cut on the BDT discriminant. The
cut is chosen by maximizing
$S_{cL}$. 

We use $S_{cL}$ to illustrate the expected improvement in sensitivity 
of the BDT method relative to the conventional one. 
A more precise estimate of the sensitivity of the LHC experiments for
observing a possible CHAMP signal would require a full detector 
simulation. However, the main point we wish to make is that a better
use of the energy depositions can enhance significantly the 
discovery potential of the LHC using the early data. The 
quantity $S_{cL}$ is sufficient to make this point.

\section{\label{sec:discussion} Application of the BDT Discriminant}

\subsection{\label{sec:discoverypotential} Discovery potential for CHAMPs }

For the two benchmark models, Table~\ref{tab:nnvscounting} compares 
the $S_{cL}$ values expected using a simple counting method with 
those expected from the BDT method.  We have assumed an integrated 
luminosity of 200 pb$^{-1}$, collected at an LHC center of mass 
energy of 10 TeV.  We note that a simple counting experiment can 
already give an $S_{cL}$ as high as 
5.6$\sigma$ (stau156 model with nominal background) using our selection of high
momentum tracks within the muon acceptance.  Even without looking at 
the energy deposition differences between muons and CHAMPs, a low mass 
CHAMP discovery is already possible for some model/mass points. 

\begin{table}
\centering \caption{Signal significance assuming 200 $pb^{-1}$ data at $\sqrt{s} = 10$ TeV LHC, using simple counting method or BDT. }
\begin{tabular}{l|c|c|c|c}
\hline
   & \multicolumn{2}{c|}{nominal} & \multicolumn{2}{c}{double nominal} \\ 
  model & \multicolumn{2}{c|}{background} & \multicolumn{2}{c}{background} \\ 
             &counting&BDT& counting & BDT  \\ 
\hline
stau156       & 5.6$\sigma$ & 8.3$\sigma$ & 4.4$\sigma$ & 7.1$\sigma$ \\ 
di-CHAMP182 & 4.2$\sigma$ & 6.7$\sigma$ & 3.2$\sigma$ & 5.8$\sigma$ \\ 
\hline
\end{tabular}
\label{tab:nnvscounting}
\end{table}

The background could be larger than our
estimate due to several effects.  The PDFs are not well constrained
at LHC energies.  Our simplified {\sc GEANT4} simulation does not
include track momentum mis-measurement that could cause moderate
momentum (100-500 GeV) muons to reconstruct at larger momentum
($>$ 500 GeV).  A full detector simulation and reconstruction as
well as comparison to real data is necessary to incorporate these
effects.  However, Table~\ref{tab:nnvscounting} also shows the 
effect of doubling the background estimate to simulate these types
of effects.  The larger background would lower the significance of
both methods.  However, the BDT discriminant can ensure discoveries 
($>$ 5$\sigma$ significance) for the stau156 and di-CHAMP182 
benchmarks, even if the background
were larger than our nominal estimate by a factor of two.

The method proposed here has very high detection efficiency 
because the CHAMP is moving very fast and therefore, like a high $p_T$ muon, its trigger and 
reconstruction efficiencies are very high at CMS and ATLAS.  (For 
$\beta$ less than $\sim$0.6, however,  the trigger and reconstruction efficiencies 
for CHAMPs drop significantly at
CMS~\cite{hscpcmspas} and ATLAS~\cite{atlashscp}.)
Moreover, the new method can use high momentum cosmic muons to evaluate 
instrumental effects related to energy depositions.  Recent results
from the CMS collaboration show the {\sc GEANT4} simulation is consistent
with detector response to cosmic 
muons~\cite{cmscraftsst, cmscraftecal, cmscrafthcal}. 

We then compare the expected significance of the low $\beta$ 
method, the high $p$ counting method, and the new BDT method for the 
two benchmark models.  Table~\ref{tab:conventionalnewcomp2} lists the 
$S_{cL}$ values from the three methods, assuming 200 pb$^{-1}$ data from 
10 TeV running at the LHC, for several mass points.  The efficiency is 
obtained from the number of CHAMP tracks in the slow ($0.6<\beta <0.8$ and 
$|\eta|<$ 0.8 and $p_T >$ 40 GeV/c) or fast 
CHAMP regions (as defined in this paper), divided by the total number 
of events.  We assume 100\% trigger and reconstruction efficiency 
for CHAMPs in both the conventional and BDT methods, 
a background of $< 1$ event for the conventional method~\cite{hscpcmspas}, 
and our nominal background estimate for the new method.

\begin{table}
 \centering \caption{Signal significance from the conventional, counting, and BDT method, assuming 200 $pb^{-1}$ data from 10 TeV LHC. The efficiency for the events to fall within the basic selection (see text) is given in parenthesis.\label{tab:conventionalnewcomp2}}
\begin{tabular}{llccc}
\hline
 \multicolumn{2}{c}{CHAMP mass} & & & \\ 
 \multicolumn{2}{c}{(GeV/$c^2$)} & ~~~Conventional~~ & ~~Counting~~ & New BDT \\ 
\hline
 \multicolumn{5}{l}{mGMSB stau}  \\ 
 ~~~ & 156 & $>$ 6.8$\sigma$ (0.20) & 5.6$\sigma$ (0.32) & 8.3$\sigma$  \\ 
 ~~~ & 200 & $>$ 3.0$\sigma$ (0.24) & 2.2$\sigma$ (0.40) & 3.9$\sigma$  \\ 
 ~~~ & 247 & $>$ 1.3$\sigma$ (0.28) & 0.8$\sigma$ (0.42) & 1.4$\sigma$  \\ 
 ~~~ & 308 & $>$ 0.5$\sigma$ (0.33) & 0.3$\sigma$ (0.49) & 0.5$\sigma$  \\ \hline
 \multicolumn{5}{l}{di-CHAMP resonance} \\ 
 ~~~ & 182 & $>$ 4.2$\sigma$ (0.19) & 4.2$\sigma$ (0.41) & 6.7$\sigma$  \\ 
 ~~~ & 242 & $>$ 2.0$\sigma$ (0.25) & 1.9$\sigma$ (0.57) & 3.3$\sigma$  \\ 
 ~~~ & 302 & $>$ 1.0$\sigma$ (0.28) & 1.0$\sigma$ (0.72) & 1.4$\sigma$  \\ 
 \hline
\end{tabular}
\end{table}

\begin{figure}
\includegraphics[width=8cm]{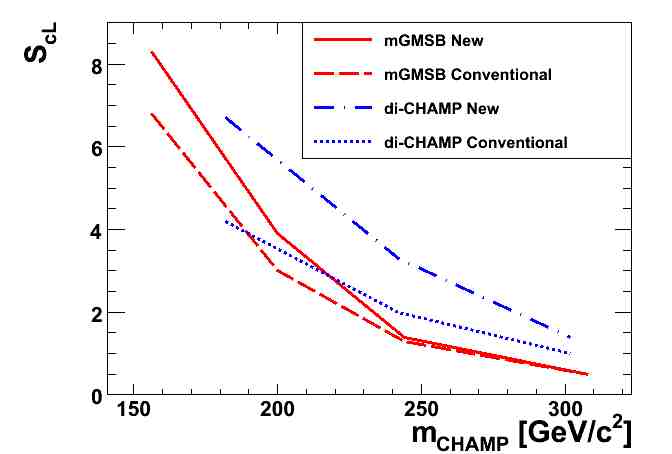}
\caption{\label{fig:sigvsmass} Significance of the new and conventional
   search methods as a function of the CHAMP mass for the two models
   explored: mGMSB SPS7 (red) and di-CHAMP (blue).}
\end{figure}

Figure~\ref{fig:sigvsmass} shows the significance of the new and
conventional methods as a function of CHAMP mass.  We see that for 
the masses of 
interest for early LHC running, the high momentum analysis has
comparable or better discovery potential.  The curve for the
conventional method is a rough estimate and it could turn out to
be larger, but we do not expect it to be radically different.
The two search methods are complementary as they explore 
different regions of signal acceptance.  Therefore, the
discovery potential could be enhanced further by combining the 
two methods, taking due account  of correlations.

Although this paper uses CMS as a model detector, the method
is applicable to ATLAS, which has similar Hcal and Ecal energy resolutions as CMS~\cite{atlasperformance}. 
Although ATLAS does not measure the Tracker dE/dx in the same way as CMS, 
the number of hits in the Transition Radiation Tracker (TRT) will give 
similar discriminating power.  This is because a high speed muon results 
in more high threshold hits, but a high speed CHAMP (still in the 
$\beta\gamma = $ 1$-$10 region) only gives about 2 hits, as pointed 
out in Ref.~\cite{AtlasRhad}.  In addition, the technique could be
extended to include information from other ATLAS or CMS detectors 
such as proposed in Ref.~\cite{EMsecondaryMuon}.

\subsection{\label{sec:dihscpresonance} Distinguishing di-$\mu$ signal from di-CHAMP or CHAMP-$\mu$ signal}

 \begin{figure}
\includegraphics[width=7cm]{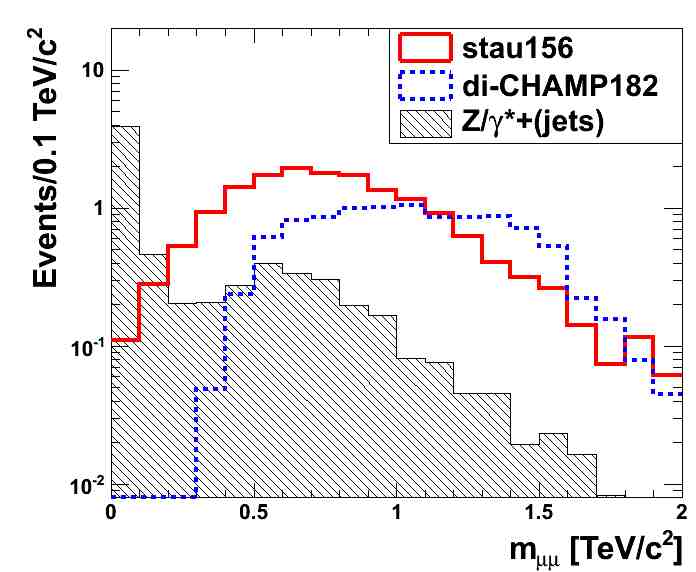}
\caption{\label{fig:dimuonmass} Di-$\mu$ mass for main background and signals (stau156 and di-CHAMP182 models, unstacked) after initial selections in this analysis. The di-$\mu$ Mass is smeared with mass resolution in Ref.~\cite{cmspasdimuon}. Plot is scaled to 200 $pb^{-1}$ 10 TeV LHC data.}
\end{figure}

In the di-CHAMP model, CHAMPs can originate from the decay
of a heavy resonance. Therefore, it is interesting to look at the
invariant mass spectrum (calculated assuming the muon mass)
for those events.  As shown in Fig.~\ref{fig:dimuonmass}, after the initial
selection used in this analysis (not including the BDT), we see a broad excess above 500 GeV/$c^2$ in the di-$\mu$
mass spectrum. Previous Monte Carlo studies have shown
some evidence that a peak may be observable~\cite{dichampph},
but it is not seen here with this benchmark model. 

Even for CHAMPs that do not originate from
a common resonance, both models predict an excess of
events at high di-$\mu$ mass (Fig.~\ref{fig:dimuonmass}). Therefore, an analysis searching the di-$\mu$ spectrum above the $Z$ peak (for evidence of a $Z^\prime$, compositeness, Randall-Sundrum gravitons,
etc.) will need to determine whether any observed excess arises from
muons or from CHAMPs.  Our new method can easily distinguish 
events with two CHAMPs from those with two muons.  The method will also be useful in
identifying CHAMP-muon resonances such as those suggested in
Ref.~\cite{leptosusy}.

\section{\label{sec:conclusion} Conclusion }

We propose a way to search for high speed CHAMPs at the LHC 
experiments to complement the conventional slowly moving CHAMP searches. 
The SM background can be kept to a very low level using simple 
selection criteria.  With 200 pb$^{-1}$ data collected at 10 TeV during the first
run of the LHC, 
a simple counting experiment already has the possibility to discover 
a stau with mass just above 140 GeV/$c^2$.  To improve the signal-background separation, 
we construct a boosted decision tree  whose inputs are the energy depositions from muon-like 
tracks.  This discriminant can improve significantly the 
discovery potential during early 
LHC running, which is our principal goal.  
By combining the two search methods, one can, in principle, increase the 
discovery potential further.  Finally, the BDT discriminant provides an 
experimental tool to distinguish CHAMP models (di-CHAMP, CHAMP-muon 
resonance models or non-resonance CHAMP models) from models with di-muon 
excesses in standard high mass di-$\mu$ searches.

\section{Acknowledgements}

This work was supported by the U.S. Department of Energy under 
grant No. DE-FG02-95ER40896.   We thank Can Kilic, Takemichi Okui and 
Raman Sundrum for providing their model code and Thomas Phillips for 
discussion of the interpretation of the CDF CHAMP search.   We also 
thank Harrison Prosper for very helpful discussions on multivariate
analysis techniques.

\bibliographystyle{aip}

\bibliography{highpchamp}

\end{document}